\documentclass[]{spie}  

\usepackage{amsmath,amsfonts,amssymb}
\usepackage{graphicx}
\usepackage[colorlinks=true, allcolors=blue]{hyperref}

\title{LitCall: {L}earning {I}mplicit {T}opology for {C}NN-based {A}ortic {L}andmark {L}ocalization}

\author[a,b]{ Zhangxing Bian}
\author[a,b]{ Jiayang Zhong}
\author[a]{Yanglong Lu}
\author[c]{ Charles R. Hatt}
\author[a]{ Nicholas S. Burris}

\affil[a]{Department of Radiology, University of Michigan, Ann Arbor, MI, USA;}
\affil[b]{Department of Electrical Engineering and Computer Science, University of Michigan, Ann Arbor, MI, USA;}
\affil[c]{Imbio LLC, Minneapolis, MN, USA;}

\authorinfo{Further author information: (Send correspondence to  Nicholas S. Burris.)\\ Nicholas S. Burris.: nburris@med.umich.edu}

 \usepackage{amsmath,amsfonts,amssymb,bm}
 \usepackage{graphicx}
 \usepackage[colorlinks=true, allcolors=blue]{hyperref}
 \usepackage[ruled,noline,linesnumbered]{algorithm2e}
 \usepackage{booktabs}
 \usepackage{multirow}
 \usepackage{siunitx}
 \usepackage{arydshln}

 \usepackage{multirow}
 \usepackage{booktabs}
 \usepackage{pifont}
 \usepackage{mathtools}
 \usepackage{authblk}
 
 \usepackage[]{caption}

 \def\BibTeX{{\rm B\kern-.05em{\sc i\kern-.025em b}\kern-.08em
 		T\kern-.1667em\lower.7ex\hbox{E}\kern-.125emX}}
 \newcommand\Tstrut{\rule{0pt}{2.6ex}}         
 \newcommand\Bstrut{\rule[-0.9ex]{0pt}{0pt}}   

\begin{document} 
\maketitle

\begin{abstract}
Landmark detection is a critical component of the image processing pipeline for automated aortic size measurements. Given that the thoracic aorta has a relatively conserved topology across the population and that a human annotator with minimal training can estimate the location of unseen landmarks from limited examples, we proposed an auxiliary learning task to learn the implicit topology of aortic landmarks through a CNN-based network. Specifically, we created a network to predict the location of missing landmarks from the visible ones by minimizing the \textit{Implicit Topology} loss in an end-to-end manner. The proposed learning task can be easily adapted and combined with Unet-style backbones. To validate our method, we utilized a dataset consisting of 207 CTAs, labeling four landmarks on each aorta. Our method outperforms the state-of-the-art Unet-style architectures (ResUnet, UnetR) in terms of localization accuracy, with only a light (\#params=0.4M) overhead.  We also demonstrate our approach in two \textit{clinically} meaningful applications: aortic sub-region division and automatic centerline generation.
\end{abstract}

\keywords{Landmark localization, deep learning, aorta, implicit topology}

\section{INTRODUCTION}
\label{sec:intro}  

Fast and accurate localization of aortic landmarks can facilitate image analysis tasks such as segmentation, registration and classification of anatomic sub-regions. In clinical practice, these tasks are performed manually, leading to significant inefficiency. Deep convolutional neural networks (CNNs) can be used to automate landmark annotation using a variety of approaches including: classifying image slices\cite{yang2015automated}, model ensembling\cite{oktay2016stratified}, landmark coordinate regression\cite{zhang2017detecting} and heatmap regression\cite{payer2019integrating}. Classifying image slices suffers from severe class imbalance, and directly regressing coordinates requires a large number of parameters and highly non-linear mapping. Heatmap regression and ensemble learning models better handle overfitting and show robustness to image variability and artifacts. Our proposed method aligns most closely with heatmap regression which assumes the probability of landmark location is not uniformly distributed over the image.

\begin{figure}[t]
	\centering
	\includegraphics[width=0.6\linewidth]{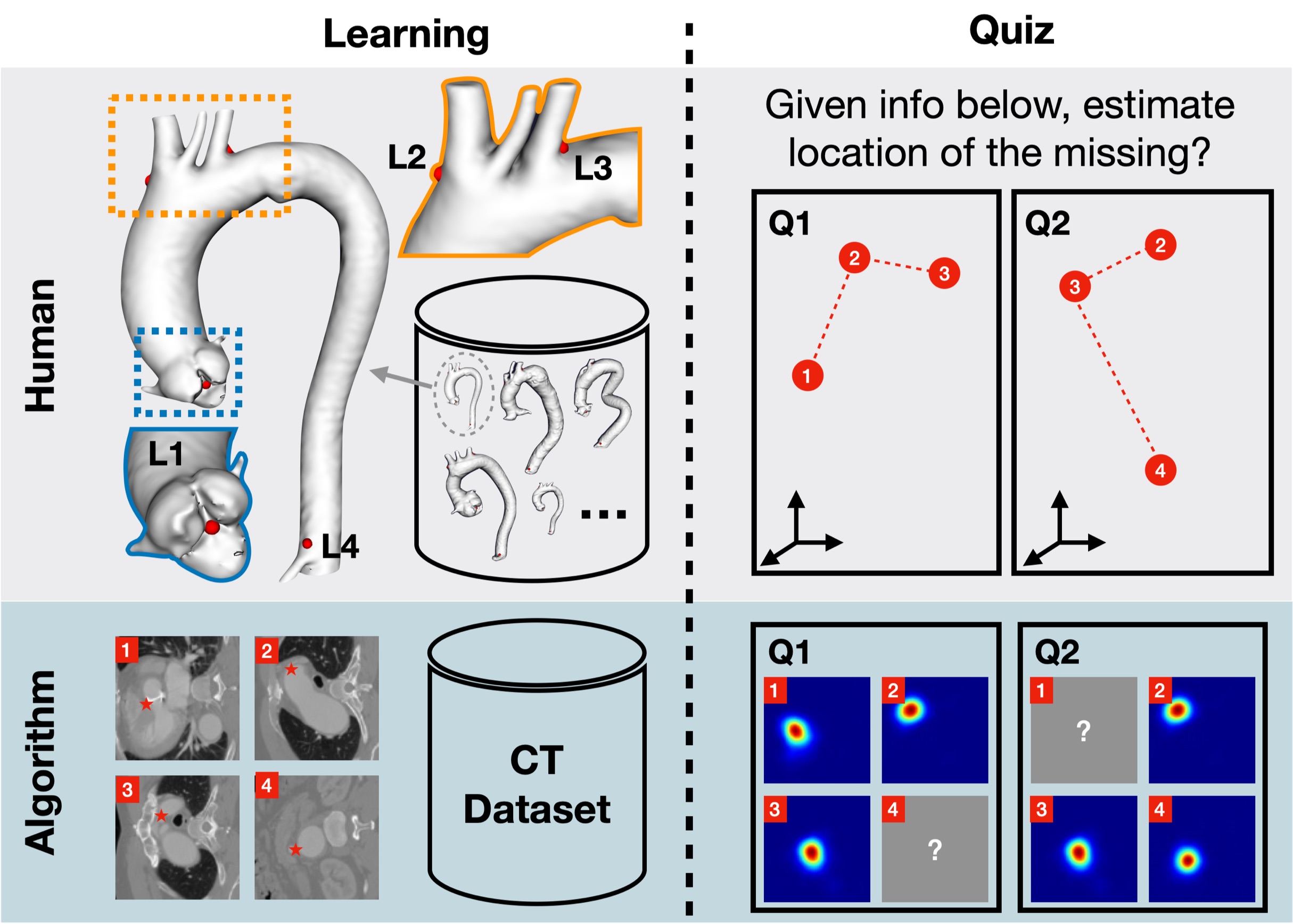}
	\vspace{0.5em}
	\caption{Humans \textit{efficiently} learn the topology of spatial patterns. Even seeing a few examples of 4 aortic landmarks, one can give a decent answer to the quizzes above. We propose an auxiliary task for a network to learn landmark topology implicitly, which greatly improves landmark localization the accuracy prevents overfitting. }
	\label{fig:teaser}
\end{figure}

Despite differences in size, tortuosity, and the location of adjacent structures between individuals, the thoracic aorta has a relatively regular shape. Thus, we propose that the topology formed by landmarks in the thoracic aorta can serve as a prior during learning. For example, a human rater can identify the approximate location of a missing landmark after observing only a few examples as shown in Fig.\ref{fig:teaser}. Inspired by this, we designed an auxiliary task for our model: learn to predict the locations of missing landmarks from a fraction of visible ones by optimizing the so-called \textit{Implicit Topology} loss in an end-to-end manner. We show that by combining the proposed learning task with Unet-style backbone, the accuracy of localization is improved with a small parameter overhead (0.4M extra parameters). 


Beyond technical implementation, we also demonstrate two clinically meaningful downstream image analysis applications: automatic centerline generation and classification of anatomic sub-regions. Generation of an aortic centerline is an important step for most two- and three-dimensional analyses of aortic disease including deformation    analysis \cite{bian2021deformable,burris2021vascular,rengier2009centerline}, however, manually labeling seeds point to generate centerline can be cumbersome.  We utilize our model to predict accurate seed locations in the ascending and distal descending aorta to automate this process. Additionally, by using the predicted landmarks as the boundary to subdivide the thoracic aorta into ascending, arch, and descending segments, this work opens the possibility of performing segment-wise assessments of aortic size and growth.

\section{METHOD}
\label{sec:method} 

Given the input CT image $I$, we predict a set of heatmaps $\mathcal{H}$ through a network parametrized by $\theta$:
\begin{equation}
	\mathcal{H} = f_h(I;\theta), \mathcal{H} = \{h_i,\dots,h_M\},
\end{equation}
where M is the number of landmarks. The final landmark coordinates are obtained as the peak responses of the predicted heatmaps.

To obtain the ground-truth heatmap, we apply a Gaussian function to the ground-truth landmarks:
\begin{equation}
	g_{i}\left(\mathbf{x} ; \sigma_{i}\right)=\frac{A}{(2 \pi)^{d / 2} \sigma_{i}^{d}} \exp \left(-\frac{\left\|\mathbf{x}-{\mathbf{L}}^*_{i}\right\|_{2}^{2}}{2 \sigma_{i}^{2}}\right)
\end{equation}

where $A$ is the scaling factor, $\mathbf{L}^*_{i}$ is the coordinate of the $i$-th landmark, and $\sigma_{i}$ controls the width of the filter response. $\bm\sigma$ is a learnable parameter that is optimized during training. To avoid a trivial solution of  $\bm\sigma$ going to infinity and network predicting a constant everywhere, we minimize $\left\|\bm\sigma\right\|_2^2$ as a extra regularization term. 

For convenience, we denote the collection of $g_i$ as set $\mathcal{G}$ and define $\mathrm{MSE}$ over two sets $\mathcal{H},\mathcal{G}$ as:

\begin{align*} 
	\mathrm{MSE}(\mathcal{H},\mathcal{G},\bm{\sigma}) = \frac{1}{MN} \sum_{\mathbf{x}}\left\|h_{i}-g_{i}\left(\mathbf{x} ; \sigma_{i}\right)\right\|_{2}^{2}, \\
	\; h_i\in\mathcal{G}, g_i \in \mathcal{H}, \mathopen|\mathcal{G}\mathclose|=\mathopen|\mathcal{H}\mathclose|
\end{align*}
where $N$ is the number of voxels, thus, the heatmap regression loss can be simply defined as:

\begin{equation}
	\mathcal{L}_\mathtt{reg} = \mathrm{MSE}(\mathcal{H},\mathcal{G};\bm{\sigma}^h)
\end{equation}

Then we introduce a selection module $s$, which randomly select $k$ maps from set $\mathcal{H}$ as visible heatmaps. $k$ will gradually increase during the training. The heatmaps of a missing landmark $\tilde{\mathcal{H}}$ can be obtained by: 
\begin{equation}
	\tilde{\mathcal{H}} = f_{\tilde{h}}(s(\mathcal{H};k);\theta, \phi), \tilde{\mathcal{H}} =\mathcal{H} - s(\mathcal{H};k), 
\end{equation}
where $f_{\tilde{h}}$ is the function parametrized by both $\theta$ and $\phi$. Note that the input for function $f_{\tilde{h}}$ only contains the predicted visible heatmap but without image features, which eliminates the trivial solution of $f_{\tilde{h}}$ imitating $f_{h}$. 

Then the \textbf{i}mplicit \textbf{t}opology loss can be simply noted as:

\begin{equation}
	\mathcal{L}_\mathtt{it} = \mathrm{MSE}(\tilde{\mathcal{H}},\tilde{\mathcal{G}}; \bm{\sigma}^{\tilde{h}}, k)
\end{equation}

With aforementioned, the overall loss function is:

\begin{equation}
	\mathcal{L} = \mathcal{L}_\mathtt{reg} + \alpha\mathcal{L}_\mathtt{it} + \beta(||\bm\sigma^{h}||_2^2 + ||\bm\sigma^{\tilde{h}}||_2^2 )
	\label{eq:loss}
\end{equation}
Note that there are two $\bm\sigma$'s in Eq.\ref{eq:loss}, one for $\mathcal{L}_\mathtt{reg}$ and another for $\mathcal{L}_\mathtt{it}$. This design decouples two learning tasks: predicting \textit{accurate} landmark location by pushing $\bm\sigma$ as small as possible and estimation of an \textit{approximate} region for the missing landmarks in a more forgiving fashion with a moderate $\bm\sigma$. 

The network is trained to minimize the loss function Eq. \ref{eq:loss} in an end-to-end manner. 

\begin{figure*}[t]
	\centering
	\includegraphics[width=\linewidth]{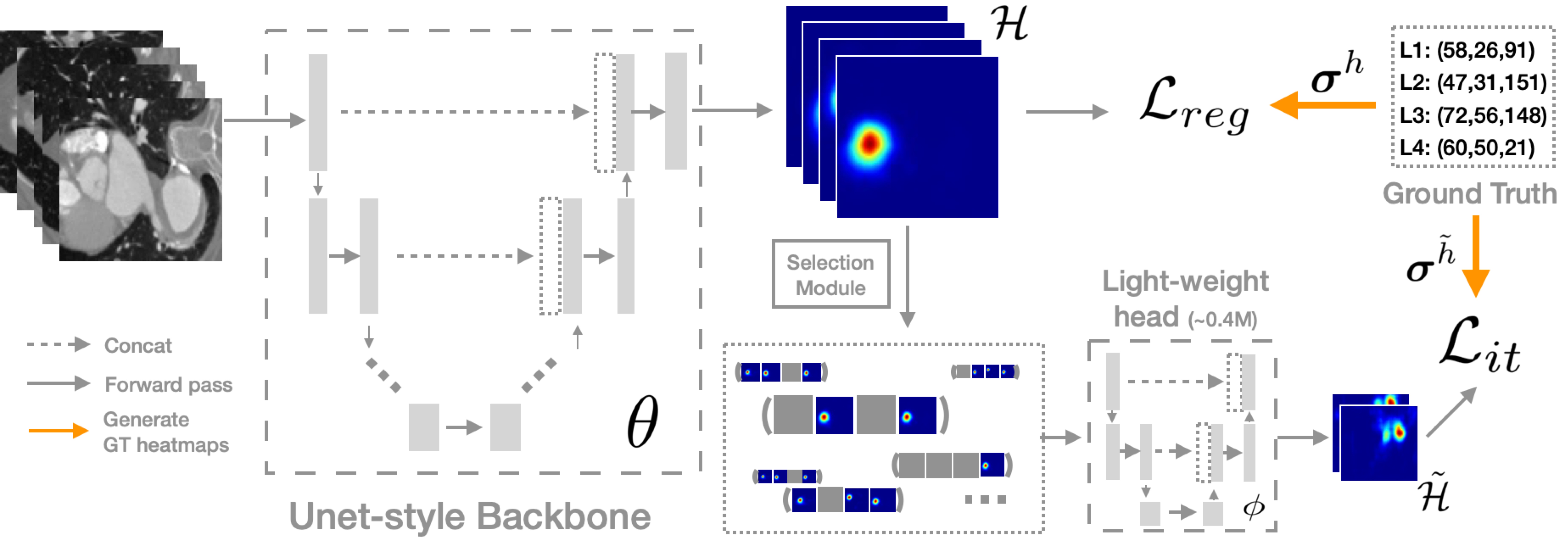}
	\caption{Illustration of our pipeline. The uncropped image is sent to a Unet-style backbone to regress heatmaps. The predicted heatmaps pass through the selection module to a light weight head, which is a three-scale resUnet, to predict the heatmaps of missing landmarks. $\mathcal{L}_{reg}$ and $\mathcal{L}_{it}$ indicate losses for regression and learning implicit topology.}
	\label{fig:method}
\end{figure*}

\section{Results}
\label{sec:exp} 

\begin{figure*}[ht]
	\centering
	\includegraphics[width=\linewidth]{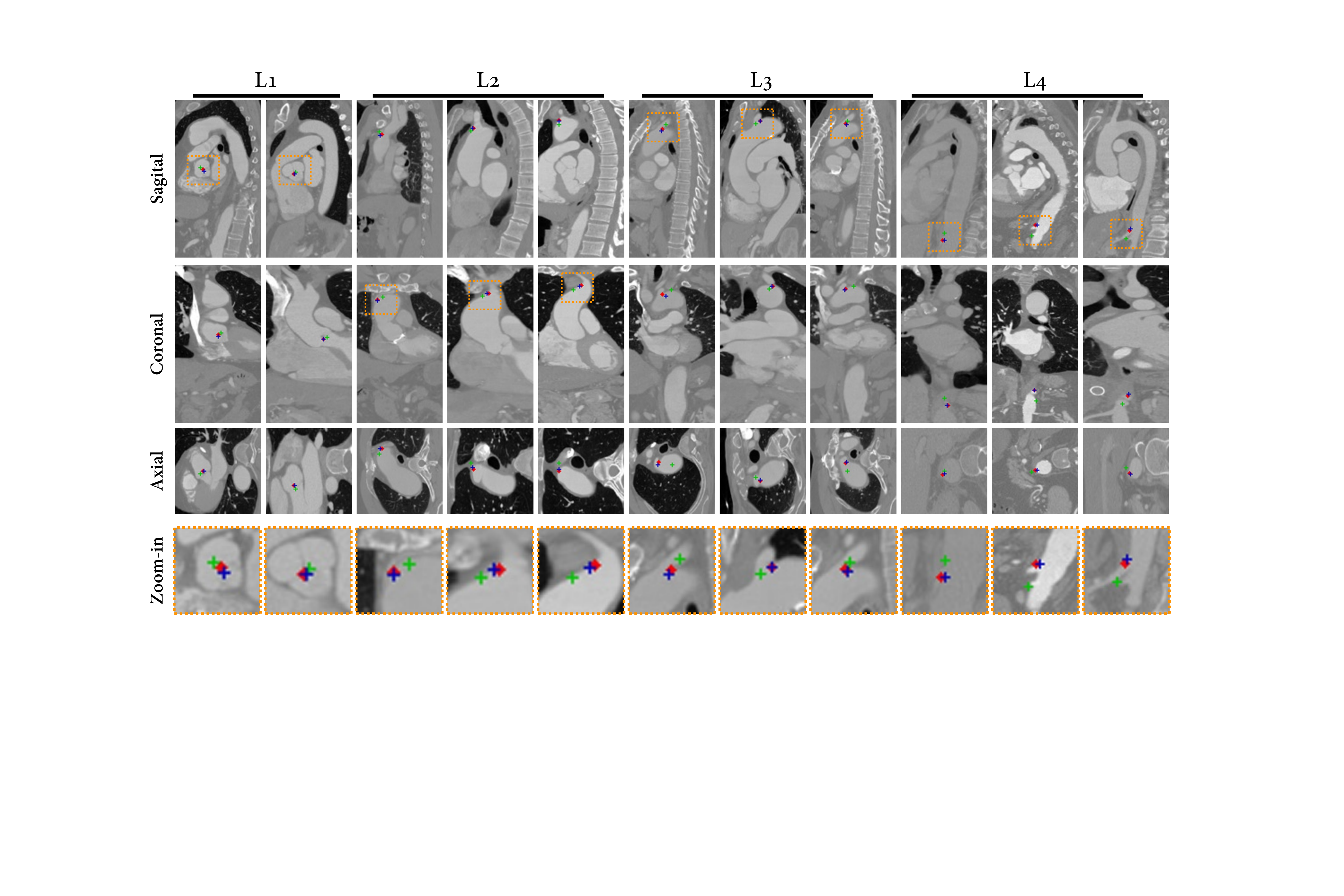}
	\caption{Qualitative result on unseen examples. For clarity, we only visualize the result of two best models, namely ResUnet+Lit and ResUnet, for comparision. {\color{red}$\blacksquare$} $\leftrightarrow$ ground truth location; {\color{blue}$\mathbf{+}$} $\leftrightarrow$ ResUnet+Lit; {\color{green}$\mathbf{+}$} $\leftrightarrow$ ResUnet. 
	}
	\label{fig:qualit}
\end{figure*}

\textbf{Dataset \& Training Details}. 
We collected 207 thoracic computed tomography angiography (CTA) scans. CTA scans were performed with electrocardiogram (ECG) gating during the administration of iodinated intravenous contrast, with images reconstructed at 75\% of the R-R interval. All images are pre-cropped from just above the aortic arch through the upper abdomen (i.e., celiac artery). This dataset includes various aortic pathologies including aneurysm and dissection in patients with and without aortic endografts. The average volume size is $230\times230\times440$ with a voxel spacing of $0.64\times0.64\times0.75 \; mm^3$. Images were resampled to have isotropic spacing of $1.8\times1.8\times1.8 \; mm^3$. We center-crop/pad input images along the three dimensions to create a $96\times96\times176$ patch, which is also the size of target heatmaps. We then clip the image intensity [-1000,1000] and normalize to [0,1]. Image data was augmented by adding Gaussian noise and random rotations during training.

For training the networks, we used the loss function in Eq. \ref{eq:loss} with parameters $\alpha = 0.1, \beta = 10^{-4}, A = 10^{6}$ that were empirically determined, with $M=4$ landmarks in total. We initialized $\bm\sigma$ with 10 voxels. Using the selection module, heatmaps were randomly erased with a probability $p$, which linearly increased from 0 to 0.5 during over the entire training procedure. We used the AdamW optimizer and CyclicLR ($base\_lr = 2\times10^{-3},max\_lr = 10^{-2}$) as a learning rate scheduler, with a mini-batch size of 2 and 20,000 total iterations. The training and inference code were implemented  in \texttt{pytorch}. Using 5-fold cross-validation, 80\% of the subjects were randomly chosen for training and the remaining were used for validation. A GTX 2080Ti GPU was used for training and inference. 

We used three Unet-like architectures: vanilla Unet~\cite{ronneberger2015u}, ResUnet~\cite{kerfoot2018left}, and UnetR~\cite{hatamizadeh2021unetr}. Most recently, UnetR combined the original Unet with a transformer module to achieve state-of-the-art performance on several 3D medical image segmentation benchmarks. Since all of these architectures are proposed to predict segmentation, i.e., pixel-wise class label, we can simply adapt these architectures as a backbone in our \textit{implicit topology learning} framework to perform landmark localization tasks, where the number of output channels is equal to the number of landmarks rather than number of label classes. The light-weight head is a smaller version of ResUnet, consisting of three scales with the number of channels (16,32,64). We use a strided convolution of $stride=4$ between each scale. We also trained larger models ("Unet-L" and "ResUnet-L") with 16 more channels for each scale to investigate the effect of increasing model parameters on performance. 

The primary evaluation metric was landmark localization accuracy, defined as the Euclidean distance between predicted landmarks and ground-truth landmarks. 

\begin{figure}
	\begin{minipage}[]{0.49\textwidth}
		\centering
		\includegraphics[width=1.0\linewidth]{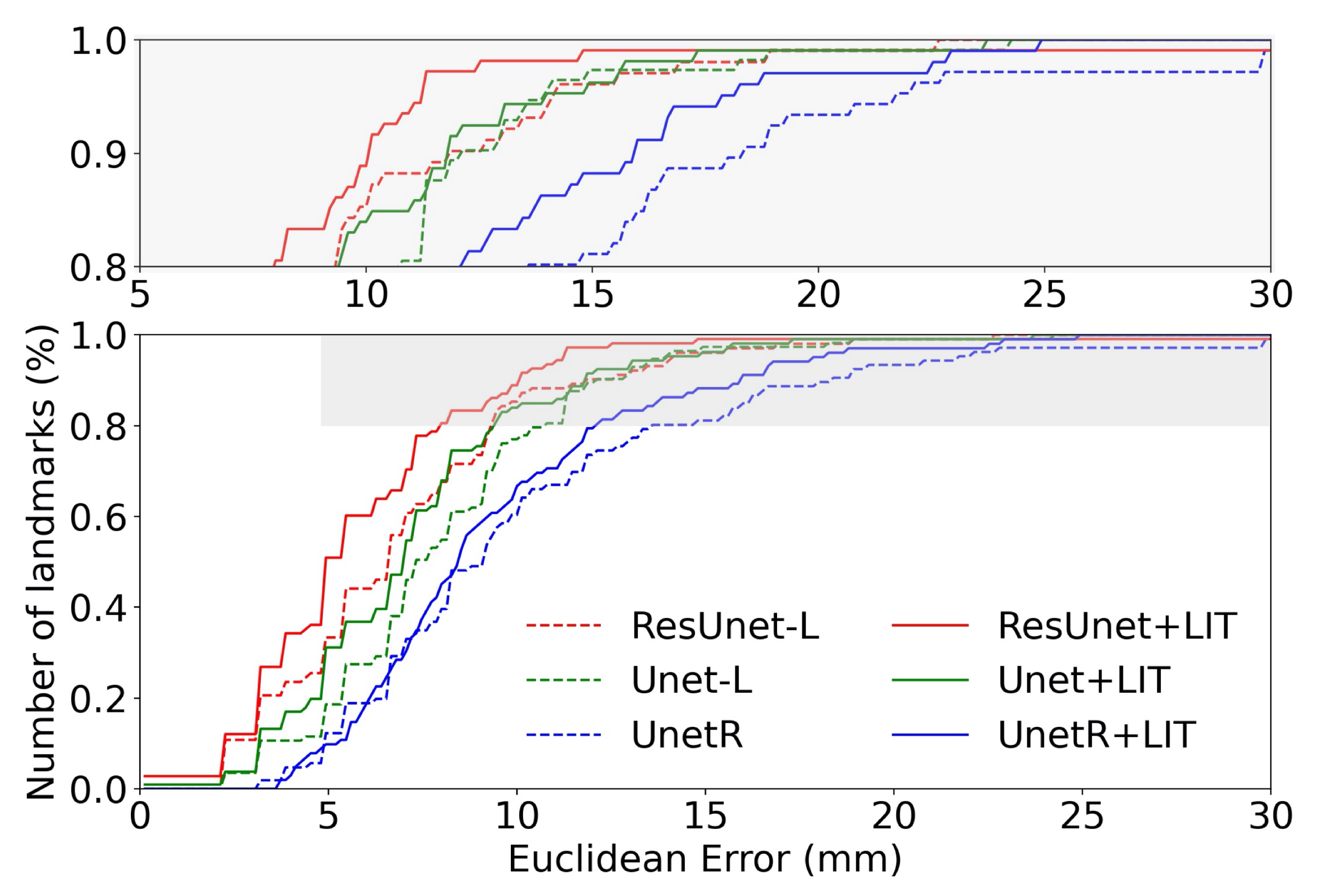} 
		\captionof{figure}{Cumulative landmark localization error. Range (0.8-1.0) is zoomed in for better visibility.}
		\label{fig:cdf}
	\end{minipage}
	\hfill
	\begin{minipage}[]{0.49\textwidth}
		\captionof{table}{Landmark localization errors. {\color{red}{Red}} is the best over column. \textbf{Bold} is the best for certain backbone. ``-L'' indicates the model with a larger number of parameters. ``LIT'' indicates the model that learns implicit topology.}
		\centering
		\resizebox{\textwidth}{!}{
		\begin{tabular}{cccc}
			\toprule
			\multirow{2}{*}{Method}      & \multicolumn{2}{l}{Euclidean Error in (mm)} & \multirow{2}{*}{\# params} \\ \cline{2-3}
			& median &           mean $\pm$ std           &                            \\ \hline
			Unet\cite{ronneberger2015u}\Tstrut\Bstrut                &  7.30  &          8.04 $\pm$ 3.57           &           2.0 M            \\
			Unet-L               &    7.28    &      7.98 $\pm$ 3.59                              &              2.7 M              \\
			Unet + \textbf{\color{blue}{LIT}} &  \textbf{6.76}  &      \textbf {7.38} $\pm$ 3.58           &           2.4 M            \\
			\cdashline{1-4}
			ResUnet\cite{kerfoot2018left}\Tstrut\Bstrut     & 6.63   &         7.01 $\pm$  3.96          &           4.8 M            \\
			ResUnet-L     & 6.60        &   6.98 $\pm$ 4.10      &                                   6.3 M                          \\
			ResUnet + \textbf{\color{blue}{LIT}}           &  \color{red}\textbf{4.92}  &          \textbf{\color{red}{5.93}} $\pm$ 4.17           &           5.2 M            \\
			\cdashline{1-4}
			UnetR\cite{hatamizadeh2021unetr}\Tstrut\Bstrut      &  9.07  &          10.40 $\pm$ 5.71          &           92.5 M           \\
			UnetR + \textbf{\color{blue}{LIT}}            &  \textbf{8.46}  &          \textbf{9.35} $\pm$ 4.27           &           92.9 M           \\ \bottomrule
		\end{tabular} \label{tab:comp}
	}
		
	\end{minipage}
\end{figure}

\textbf{Quantitative \& Qualitative  Results}. 
The mean and SD of the landmark localization error is reported in Table \ref{tab:comp}. Cumulative landmark error distributions in Fig.\ref{fig:cdf}. Qualitative examples are illustrated in in Fig.\ref{fig:qualit}. It can be observed in Table.\ref{tab:comp} that UnetR suffered from overfitting due to the larger model capacity. Interestingly, when combining UnetR with the implicit topology learning task, the overfitting problem is alleviated and performance improves by 10\% with only a 0.4\% increase of parameters. 


\begin{figure}[tbp]
	\vspace{1em}
	\centering
	\includegraphics[width=0.7\linewidth]{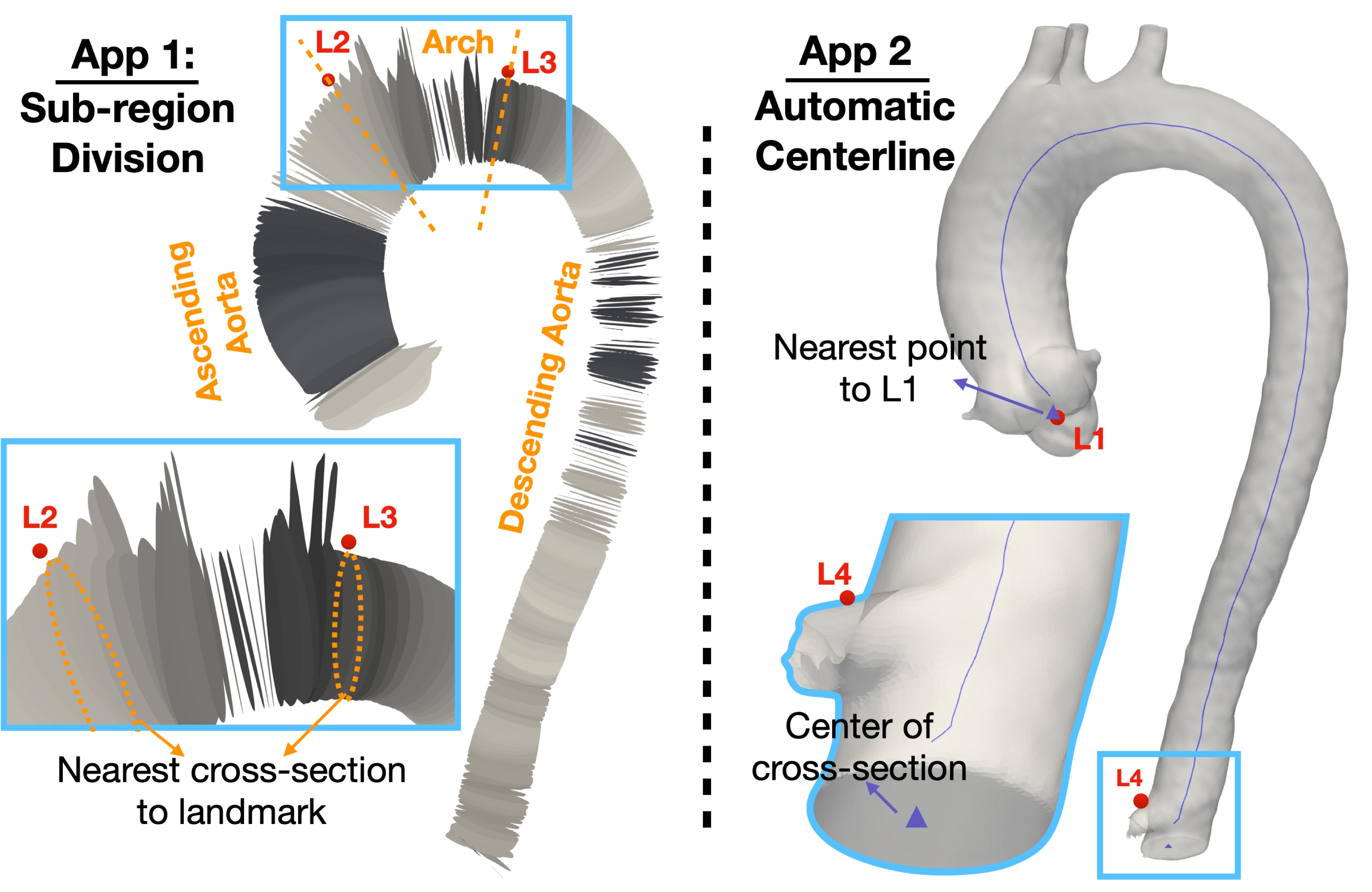}
	\caption{Two applications of aortic landmark localization: sub-region division and automatic centerline generation. When applied to divide the aorta into sub-regions, L2 and L3 are used to determine the boundary cross-sections which divides the aorta into three parts: ascending, arch, and descending  aorta (App 1).  When applied to automatic centerline generation,  L1 and L4 are used  to determine  the  start and end seed for generating the centerline (App 2). }
	\label{fig:app}
\end{figure}

\textbf{Clinical Applications}. Aortic centerline generation is a necessary step for most clinical and research analyses of aortic geometry. However, the vast majority of centerline algorithms require manual interaction (e.g. placement of starting and ending seed-points). Based on our method, we show this task can be fully automated without human intervention.

To automate centerline generation, we ran our previously developed aorta segmentation network\cite{zhong2021segmentation} to obtain an aorta mask, and then applied the method developed in this work to obtain the L1 (ascending root) and L4 (descending celiac) landmarks, which are taken as the seeds by Vascular Modeling Toolkit (VMTK) to reconstruct the 3D mesh and centerline (Figure \ref{fig:app}). This pipeline is fully implemented in python.

Recent work by our group has focused on developing techniques to quantify aortic growth/deformation over time using B-spline-based deformable registration methods\cite{bian2021deformable, burris2021vascular} to analyze aortic aneurysms. Currently, these techniques consider the whole aorta to compute deformation metrics and statistics. Equipped with the landmark detection method described here, one can automatically subdivide analysis of the entire thoracic aorta into three sub-regions, ascending aorta, arch, descending aorta, and descending root. However, the utility of defining aortic sub-regions is not limited to such novel registration-based growth assessment techniques, but can also be used to yield regional assessments of conventional metrics of aortic disease such as maximal diameter and volume.  These unique aortic segments are affected  differently by disease and thus the ability to analyze each separately may better allow regional assessments of disease.

\section{CONCLUSION}
\label{sec:conclusion} 

We proposed a simple yet effective learning task to make the network learn the implicit topology of the thoracic aorta. The proposed method can be easily combined with Unet-style backbones and is trainable in an end-to-end manner. Localization accuracy is  improved compared to baseline and the overfitting problem is alleviated. We believe this method is broadly applicable, and in future work we plan to apply our method to anatomic landmark localization tasks in other anatomies (e.g., hand joints, spine). 

\bibliography{main} 

\begin{thebibliography}{10}

\bibitem{yang2015automated}
Yang, D., Zhang, S., Yan, Z., Tan, C., Li, K., and Metaxas, D., ``Automated
  anatomical landmark detection ondistal femur surface using convolutional
  neural network,'' in [{\em 2015 IEEE 12th international symposium on
  biomedical imaging (ISBI)}{\nolinebreak\hspace{0.1em}]},   17--21, IEEE
  (2015).

\bibitem{oktay2016stratified}
Oktay, O., Bai, W., Guerrero, R., Rajchl, M., de~Marvao, A., O’Regan, D.~P.,
  Cook, S.~A., Heinrich, M.~P., Glocker, B., and Rueckert, D., ``Stratified
  decision forests for accurate anatomical landmark localization in cardiac
  images,'' {\em IEEE transactions on medical imaging}~{\bf 36}(1),  332--342
  (2016).

\bibitem{zhang2017detecting}
Zhang, J., Liu, M., and Shen, D., ``Detecting anatomical landmarks from limited
  medical imaging data using two-stage task-oriented deep neural networks,''
  {\em IEEE Transactions on Image Processing}~{\bf 26}(10),  4753--4764 (2017).

\bibitem{payer2019integrating}
Payer, C., {\v{S}}tern, D., Bischof, H., and Urschler, M., ``Integrating
  spatial configuration into heatmap regression based cnns for landmark
  localization,'' {\em Medical image analysis}~{\bf 54},  207--219 (2019).

\bibitem{bian2021deformable}
Bian, Z., Zhong, J., Hatt, C.~R., and Burris, N.~S., ``A deformable image
  registration based method to assess directionality of thoracic aortic
  aneurysm growth,'' in [{\em Medical Imaging 2021: Image
  Processing}{\nolinebreak\hspace{0.1em}]},   {\bf 11596},  115962P,
  International Society for Optics and Photonics (2021).

\bibitem{burris2021vascular}
Burris, N., Bian, Z., Dominic, J., Zhong, J., Van~Bakel, D., Houben, I., Patel,
  H., Ross, B., Christensen, G., and Hatt, C., ``Vascular deformation mapping
  as a method for 3d growth mapping during ct surveillance of thoracic aortic
  aneurysm,'' {\em Radiology}  (doi: http://dx.doi.org/10.7302/246, 2021).

\bibitem{rengier2009centerline}
Rengier, F., Weber, T.~F., Giesel, F.~L., Bockler, D., Kauczor, H.-U., and von
  Tengg-Kobligk, H., ``Centerline analysis of aortic ct angiographic
  examinations: benefits and limitations,'' {\em American Journal of
  Roentgenology}~{\bf 192}(5),  W255--W263 (2009).

\bibitem{ronneberger2015u}
Ronneberger, O., Fischer, P., and Brox, T., ``U-net: Convolutional networks for
  biomedical image segmentation,'' in [{\em International Conference on Medical
  image computing and computer-assisted
  intervention}{\nolinebreak\hspace{0.1em}]},   234--241, Springer (2015).

\bibitem{kerfoot2018left}
Kerfoot, E., Clough, J., Oksuz, I., Lee, J., King, A.~P., and Schnabel, J.~A.,
  ``Left-ventricle quantification using residual u-net,'' in [{\em
  International Workshop on Statistical Atlases and Computational Models of the
  Heart}{\nolinebreak\hspace{0.1em}]},   371--380, Springer (2018).

\bibitem{hatamizadeh2021unetr}
Hatamizadeh, A., Yang, D., Roth, H., and Xu, D., ``Unetr: Transformers for 3d
  medical image segmentation,'' {\em arXiv preprint arXiv:2103.10504}  (2021).

\bibitem{zhong2021segmentation}
Zhong, J., Bian, Z., Hatt, C.~R., and Burris, N.~S., ``Segmentation of the
  thoracic aorta using an attention-gated u-net,'' in [{\em Medical Imaging
  2021: Computer-Aided Diagnosis}{\nolinebreak\hspace{0.1em}]},   {\bf 11597},
  115970M, International Society for Optics and Photonics (2021).

\end{thebibliography}
\bibliographystyle{spiebib} 

\end{document}